\documentstyle[editedvolume,numreferences]{crckapb}

\renewcommand{\vec}[1]{{\bf #1}}

\newcommand{\be}{\begin{equation}}
\newcommand{\ee}{\end{equation}}
\newcommand{\bea}{\begin{eqnarray}}
\newcommand{\eea}{\end{eqnarray}}
\newcommand{\la}{\langle}
\newcommand{\ra}{\rangle}
\newcommand{\lb}{\left[}
\newcommand{\rb}{\right]}
\newcommand{\lp}{\left(}
\newcommand{\rp}{\right)}

\renewcommand{\H}{{\cal H}}
\newcommand{\U}{{\cal U}}

\begin{opening}
\title{THE STATISTICAL THEORY OF MESOSCOPIC NOISE}
\subtitle{A short review}

\author{L.S.\,Levitov}
\institute{Department of Physics, 
Massachusetts Institute of Technology, 77 Massachusetts Ave, 
Cambridge, MA 02139}

\end{opening}

\runningtitle{COUNTING STATISTICS OF MESOSCOPIC NOISE}

\begin{document}

\begin{abstract}
Microscopic theory of counting statistics of electrical noise 
is reviewed. We discuss a model of passive charge detector 
based on current fluctuations coupled to a spin, and its relation 
with the theory of photon counting in quantum optics. 
The statistics of tunneling current and, in particular, the 
properties of the third moment are studied in detail. 
The third moment is shown to be temperature-independent
for tunneling in a generic many-body system. Then the statistics 
of mesoscopic transport is discussed. We consider applications of 
the functional determinant formula for the generating function
of counting distribution to the DC and photo-assisted transport, 
and to mesoscopic pumping. A universal dependence 
of the noise in a mesoscopic pump on the pumping fields is obtained
and shown to provide a method of measuring the quasiparticle charge 
in an open system without any fitting parameters.\\
{\it Key words:} 
counting statistics, third moment, photo-assisted transport, 
mesoscopic pumping
\end{abstract}


\section{Introduction}

The measurements performed by optical detectors, such as photon counters, 
are extended in the time domain, which makes them sensitive to temporal 
correlations of photons\,\cite{Mandel}. 
It has been known long ago in the theory of 
photodetection\,\cite{Glauber} that 
understanding photon counting  
is essentially a problem of many-particle statistics. 
Similar considerations apply to 
the electrical noise measurement, although it differs 
from photodetection in that the electrons, unlike photons, are not destroyed 
in the process of counting.
The noise measurement, very much like photodetection, 
is a sensitive probe of temporal correlations between 
electrons.

Fermi correlations in the electron noise were originally studied
by Lesovik\,\cite{lesovik89} (see also Ref.\,\cite{khlus87}) 
in a point contact, 
and then by B\"uttiker\,\cite{buttiker90} 
in multiterminal systems, 
and by Beenakker and B\"uttiker\,\cite{beenakker92} in mesoscopic conductors. 
Kane and Fisher proposed to employ the shot noise for detecting
fractional quasiparticles in a Quantum Hall edge system
\,\cite{kanefisher93}. Subsequent theoretical developments 
are summarized in a recent review\,\cite{blanter00}.

Experimental studies of the shot noise, after first measurements
in a point contact by Reznikov et al.\,\cite{reznikov95} 
and Kumar et al.\,\cite{kumar96},
focused on the quantum Hall regime. The fractional charges $e/3$ and $e/5$
were observed\,\cite{depicciotto97,glattli97,reznikov99} at
incompressible Landau level filling (see also recent
work on noise at intermediate filling\,\cite{heiblum}).
The shot noise in a mesoscopic conductor was observed by
Steinbach et al.\,\cite{steinbach96} and Schoelkopf et al.\,\cite{schoelkopf97},
who also studied noise in photo-assisted
phase-coherent mesoscopic transport\,\cite{schoelkopf98}.

In this article we discuss counting statistics of electric noise
and consider the probability distribution of charge transmitted
in a fixed time interval\,\cite{levitov93,ivanov93}. 
This distribution provides detailed information about current fluctuations.
The counting statistics have been studied 
for the DC transport of free fermions
\,\cite{levitov93,levitov94}, 
the photo-assisted transport\,\cite{ivanov97}, 
the parametrically driven transport\,\cite{ivanov93,levitov01'1}, 
and in the mesoscopic regime\,\cite{hwlee95}
(also, see a review\,\cite{levitov96}). 
Nazarov developed Keldysh formalism for 
the counting statistics problem and applied it to mesoscopic transport 
in a weak localization regime\,\cite{nazarov} and, together with Bagrets, 
in a multiterminal geometry\,\cite{nazarov-circuit}. Charge doubling
due to Andreev scattering in NS junctions was considered
by Muzykantskii and Khmelnitskii\,\cite{muzykantskii},
and in mesoscopic NS systems by Belzig and Nazarov\,\cite{nazarovNS}. 
Andreev and Kamenev\,\cite{AndreevKamenev} studied
the problem of mesoscopic pumping in view of the results
of Ref.\,\cite{ivanov93}.
Taddei and Fazio discussed counting statistics of entangled electron
sources\,\cite{fazio}. Statistics of transport in a Coulomb blockade 
regime was studied by Bagrets and Nazarov\,\cite{bagrets02}.
Photon statistics was considered 
by Beenakker and Schomerus\,\cite{beenakker-photons} 
and Kindermann et al.\,\cite{kindermann02}.
The problem of back influence of a charge detector on current fluctuations
in the context of counting statistics measurement was studied
by Nazarov and Kindermann\,\cite{nazarov-general}.

The possibility of measuring counting statistics using 
a fast charge integrator scheme was considered recently\,\cite{levitov01'2}. 
From the measured distribution all moments of
charge fluctuations can be calculated
and, conversely, the knowledge of all moments is in principle
sufficient for recovering the full distribution.
However, due to the central limit theorem,
high moments are probably difficult to access experimentally.
Therefore recent literature focused primarily on the third moment.
It was found that the third moment
obeys a generalized Schottky relation which holds 
in the tunneling regime at both high and low temperature, 
but involves a temperature-dependent Fano factor in the mesoscopic 
regime\,\cite{levitov01'1,gutman02}. Gutman and Gefen\,\cite{gutman02} 
studied the third moment using a sigma model approach, 
while Nagaev\,\cite{nagaev02} demonstrated that 
all moments are correctly reproduced by an extension of 
the Boltzmann-Langevin kinetic equation.

In this article, after introducing the 
counting distribution 
(Sec.\ref{sec:general}), we review its microscopic definition based on 
a passive charge detector (Sec.\ref{sec:micro-spin1/2}). 
In Sec.\ref{sec:tunneling} we study 
the statistics of tunneling in a generic many-body system. 
From the microscopic approach of Sec.\ref{sec:micro-spin1/2} we derive
a bidirectional Poisson distribution for tunneling current, obtain 
a Schottky-like relation for the third moment and discuss its robustness.
After that we discuss the relation of the counting statistics
theory and the theory of photo-detection (Sec.\ref{sec:photo-detection}). 
Then we proceed to the problem of mesoscopic transport.
In Sec.\ref{sec:DC} we review the results on the DC transport 
and the derivation of a functional determinant formula for the 
counting distribution generating function. In Sec.\ref{sec:cases}
we review the work on the AC transport statistics, and then 
consider the problem of mesoscopic pumping (Sec.\ref{sec:mesopump}). 
The counting statistics for generic pumping strategy at weak pumping 
is given by a bidirectional Poisson distribution. We show that the 
Fano factor varies between $0$ and $1$ as a function of the 
pumping fields phase difference.

\section{General approach}
\label{sec:general}

The transmitted charge distribution 
can be characterized\,\cite{levitov93,ivanov93} by electron counting probabilities 
$p_n$, usually accumulated in a generating function\footnotemark{}
\footnotetext{The function $\chi(\lambda)$ is also called 
a characteristic function\,\cite{Mandel}.}
\be\label{eq:chi-p_n}
\chi(\lambda)=\sum e^{in\lambda}\,p_n\,.
\ee
The function $\chi(\lambda)$ is $2\pi$-periodic in the {\it counting field}
$\lambda$ and has the property 
$\chi(0)=1$ which follows from the probability normalization
$\sum_n p_n=1$. The term ``counting field'' will be motivated in
Sec.\ref{sec:micro-spin1/2}, where a microscopic definition 
of $\chi(\lambda)$ is discussed in which $\lambda$ appears as
a field that couples current fluctuations to a charge detector.

The generating function (\ref{eq:chi-p_n}) is particularly well suited for 
characterizing statistics of the distribution $p_n$.
The so-called {\it irreducible correlators} 
$\langle\!\langle\delta n^k\rangle\!\rangle $ 
(also known as {\it cummulants}) are
expressed
in terms of $\chi(\lambda)$ as
  \be \label{eq:ln-chi}
\ln\chi(\lambda)=
\sum_{k=1}^{\infty} m_k\frac{(i\lambda)^k}{k!}
\, ,\qquad
m_k \equiv \langle\!\langle\delta n^k\rangle\!\rangle \,.
  \ee
The first two correlators 
in (\ref{eq:ln-chi}) give the 
mean and the variance:
\be
m_1=\overline {n},
\qquad
m_2=\overline {\delta n^2} = \overline {n^2} - \overline {n}^2 ,
\ee
where $\overline {f(n)}$ stands for $\sum_n f(n) p_n$.
The third correlator\footnotemark{}
\be\label{eq:q3ave}
m_3=\langle\!\langle\delta n^3\rangle\!\rangle
 \equiv \overline {\delta n^3}=\overline{\lp n-\overline{n}\rp^3}
\ee
characterizes the asymmetry (or skewness) of the distribution $p_n$.

\footnotetext{The relation between cummulants and correlators is generally 
more complicated than Eq.(\ref{eq:q3ave}) for the third cummulant.
For example, 
$m_4=\langle\!\langle\delta q^4\rangle\!\rangle
= \overline {\delta q^4}-3\lp \overline {\delta q^2} \rp^2$.}

To illustrate the notion of a generating function, 
let us consider a Poisson process. 
It describes charge transport at very low 
transmission, with uncorrelated transmission events.
For the Poisson distribution
\be\label{eq:chi-Poisson}
p_k=\cases{e^{-\bar n}\,\bar n^k/k! & $k\ge0$\cr 0 & $k<0$}
\quad {\rm and}\quad 
\chi(\lambda)=\exp\lp (e^{i\lambda}-1) \bar n\rp\,,
\ee
where $\bar n=It/e$ is the average number of particles transmitted during 
time $t$, with $I$ the time-averaged current and $e$
the elementary charge. Comparing (\ref{eq:chi-Poisson}) with
(\ref{eq:ln-chi}), one finds that all cummulants of the Poisson distribution
are identical: $m_k=\bar n$. 

Another useful example is binomial statistics. A binomial distribution arises 
when a fixed number $N$ of independent attempts to transmit particles is made, 
each attempt successful or unsuccessful with probabilities $p$ and $q=1-p$.
The probability to transmit $k$ particles in this case is determined by the
combinatorial number $C_N^k=N!/(N-k)!k!$ of $k$ successful outcomes.
The 
probability distribution and the generating function in this case are
\be\label{eq:chi-binomial}
p_k=C_N^k p^k q^{N-k}
\quad {\rm and}\quad
\chi(\lambda)=\lp p e^{i\lambda}+q\rp^N .
\ee
The cummulants of the binomial distribution (\ref{eq:chi-binomial}) 
can be found from (\ref{eq:ln-chi}) by expanding $\ln\chi$ in $\lambda$:
\be
m_1=pN,\quad m_2=pqN,\quad m_3=pq(q-p)N, \quad ...
\ee
The binomial distribution (\ref{eq:chi-binomial}) describes 
counting distribution of 
DC current noise for a single channel scatterer, 
such as point contact, at zero temperature (Sec.\ref{sec:DC}).

Statistically independent processes result in a generating function
given by a product of generating functions for constituting processes:
$\chi(\lambda)=\chi_1(\lambda)...\chi_k(\lambda)$. 
For example, consider {\it a biderectional
Poisson distribution} defined as a mixture of two independent 
Poisson processes transmitting particles in opposite directions
with the rates $\bar n$ and $\bar n'$. In this case,
\be\label{eq:chi-bidirectional}
\chi_{2P}(\lambda)=\exp\lp (e^{i\lambda}-1) \bar n\rp
\cdot\exp\lp (e^{-i\lambda}-1) \bar n'\rp\,.
\ee
In Sec.\ref{sec:tunneling} we use the distribution 
(\ref{eq:chi-bidirectional}) to describe 
statistics of tunneling current. In Sec.\ref{sec:mesopump} 
we show that it describes noise a mesoscopic pump.

\subsection{A microscopic representation of $\chi(\lambda)$}
\label{sec:micro-spin1/2}

Here we discuss a microscopic definition of counting statistics 
for a physical system. 
Adopting an inductive approach, we shall start with a specific 
model of current detector\,\cite{levitov94,levitov96}.
We obtain the generating function $\chi(\lambda)$ for 
a particular current measurement scheme, and then argue that it
describes generic measurement.

In a realistic noise measurement, e.g. in a mesoscopic wire or a point contact,
the current fluctuations are not detected directly. 
Instead, the measurement is performed on the electromagnetic fluctuations 
(basically, voltage noise) induced by current fluctuations in the system.
The electromagnetic fluctuations have to be amplified before being detected.
The conversion of underlying microscopic fluctuations due to fermions
(electrons, fractional charges, etc.)
into fluctuations of bosons (photons) is crucial, since Bose fields
can be amplified without compromising noise statistics, 
while Fermi statistics is not consistent with amplification.

Our goal is to clarify the microscopic picture of current fluctuations, 
rather than to describe realistic measurements.
Thus we choose a {\it gedanken} measurement scheme well suited 
for that purpose. Consider a spin $1/2$ placed near an 
electron system and magnetically coupled to the electric current. 
We restrict the coupling to
the spin $z$ component, so that the system in the presence 
of the spin is described by $\H(\vec q,\vec p-\vec a \sigma_3)$,
where $\H(\vec q,\vec p)$ is the electron Hamiltonian and 
$\vec a(r)$ is the spin vector potential scaled by $e/c$.

The scheme of current detection using such spin dynamics 
can be motivated quasiclassically. A spin coupled to 
a time-dependent classical current $I(t)$ by the interaction
$\H=\frac12\lambda\sigma_3 I(t)$ will precess at the 
rate proportional to current, which turns the spin 
into an analog galvanometer. Indeed, if the spin-current coupling
is turned on at $t=0$, the spin will start precessing
around the $z$ axis with the precession angle 
$\theta(t)=\lambda\int_0^t I(t')dt'$ equal to the 
transmitted charge times $\lambda$. The coupling constant $\lambda$, so far arbitrary, 
will be associated with counting field below.

In a fully quantum-mechanical problem, the spin evolution can be 
obtained from $i\dot\vec\sigma = \lb \vec\sigma,\H\rb$. 
Since the spin-current coupling Hamiltonian commutes with $\sigma_3$,
the spin dynamics can be found explicitly. 
For that we consider the transverse spin components
$\sigma_\pm\equiv\sigma_1\pm i\sigma_2$ and write
the evolution equation $i\dot\sigma_\pm=[\sigma_\pm,\H]$ in the form
\bea \label{eq:s+}
&& i\dot\sigma_+ = 
\sigma_+ P_{\downarrow}\H(\vec q,\vec p+\vec a )
-\H(\vec q,\vec p-\vec a )P_{\uparrow}\sigma_+ 
\,,\quad \\
\label{eq:s-}
&& i\dot\sigma_- =
\sigma_-P_{\uparrow}\H(\vec q,\vec p-\vec a ) -
\H(\vec q,\vec p+\vec a )P_{\downarrow}\sigma_- 
\,,
\eea
with $P_{\uparrow,\downarrow}=\frac12(1+\sigma_3)$ 
the up and down spin projectors.
Here we used the raising/lowering properties of the 
operators $\sigma_\pm$ and replaced
$\sigma_3$ by $1$ to the left of $\sigma_+$ and by $-1$ to the right of 
$\sigma_+$ (and similarly for the $\sigma_-$ equation). 

We consider a measurement which is performed during time interval $0<t'<t$,
i.e. start with a free spin, $\vec a_{t<0}=0$, 
couple it to the electron system at $t=0$, 
maintain a finite coupling during $0<t'<t$, and then turn it off.
The expectation value of the transverse spin component 
at the time $t$, found by integrating Eqs.\,(\ref{eq:s+}),(\ref{eq:s-}), is
\be\label{eq:s+ave}
\la\sigma_+(t)\ra \,=\, 
\la e^{i\H(\vec q,\vec p-\vec a )t}e^{-i\H(\vec q,\vec p+\vec a )t}\ra_{\rm el}
\,\la\sigma_+(0)\ra_{\rm spin},
\ee
while $\la\sigma_-(t)\ra =\la\sigma_+(t)\ra^\ast$.
Note that the result of the coupled spin and current evolution 
factors into a product of quantities that depend separately 
on electron dynamics and on the initial state of the spin, 
as indicated by the subscripts.

The effect of current on spin precession is described by the
dependence of the first term in Eq.(\ref{eq:s+ave}) 
on the gauge field $\vec a$.
To make contact with the quasiclassical discussion above, let us expand 
$\H$ in $\vec a$, assuming it to be small. The result is 
$\H(\vec q,\vec p\pm\vec a )=\H(\vec q,\vec p)\pm \vec a\vec j$,
where $\vec j$ is electric current. Substituting this back in
Eq.(\ref{eq:s+ave}) and passing to the interaction representation with 
respect to the Hamiltonian of fermions uncoupled from the spin,
we rewrite the average $\la ...\ra_{\rm el}$ in (\ref{eq:s+ave}) as
\be\label{eq:factorTT}
\left\langle {\rm {\widetilde T}exp\,}\lp i\int_t^0 \int\vec a\vec j(t')d^3 r\, dt'\rp
{\rm Texp\,}\lp -i\int_0^t \int\vec a\vec j(t')d^3 r\,dt'\rp 
\right\rangle_{\rm el}\,.
\ee
Let us consider
a specific form of $\vec a$, taking it to be a pure gauge, 
$\oint \vec a d\vec l=0$,
within the electron system, and nonzero near a particular surface
(e.g. a $\delta-$function on the surface).
For a classical current, ignoring noncommutativity of current operators
at different times, the 
expression (\ref{eq:factorTT}) becomes
%
\be\label{eq:factor-classical}
\left\langle e^{ -i\theta(t)} \right\rangle_{\rm el}
\quad {\rm with} \quad
\theta(t)=\lambda \int_0^t I(t') dt' .
\ee
Here $I(t)$ is the total current through the surface
and $\lambda=-2\int \vec a d\vec l$, where the integral is taken 
across the surface. The form of Eq.(\ref{eq:factor-classical})
agrees with what one expects for 
the precession phase factor averaged over classical current fluctuations. 
For $n$ electrons transmitted through the system during 
the measurement time, the precession angle is $\theta(t)=\lambda n$. 
This relates the average in (\ref{eq:factor-classical}) 
with the transmitted charge distribution:
\be\label{eq:class-phase-ave}
\left\langle e^{ -i\theta(t)} \right\rangle_{\rm el}
=\sum_n e^{i\lambda n} p_n\,.
\ee
The relation with the spin precession in this case can be seen more clearly 
by combining the result (\ref{eq:class-phase-ave}) with
Eq.(\ref{eq:s+ave}), 
\be\label{eq:class-s+ave}
\la\sigma_+(t)\ra = \sum_n  p_n\,\lp e^{i\lambda n} 
\la\sigma_+(0)\ra_{\rm spin}\rp\,,
\ee
and recalling the transformation rule $\sigma_+'=e^{i\theta}\sigma_+$
for spin rotation around the $z$-axis by an angle $\theta$. 
This way of writing the result of spin evolution confirms the
expected relationship between the charge counting probability 
distribution and the distribution of the spin precession angles.

This discussion clarifies the meaning of the quantity $\la ...\ra_{\rm el}$
in Eq.(\ref{eq:s+ave}), linking it to the counting distribution 
generating function. Motivated by this, we use Eq.(\ref{eq:s+ave}) to give 
a microscopic definition of
counting statistics. We rewrite the quantity $\la ...\ra_{\rm el}$
as a Keldysh partition function
  \be\label{eq:Zkeldysh}
\chi(\lambda) = \Big\langle {\rm T}_{\rm K} 
\exp\lp -i\int_{C_{0,t}}\hat{\H}_{\lambda}(t')dt'\rp\Big\rangle_{\rm el}
  \ee
with the integral taken over the Keldysh time contour $C_{0,t}\equiv[0\to t\to 0]$, 
first forward and then backward in time. The counting field $\lambda$ 
is related to the gauge field $\vec a$ via 
$\lambda=\mp\frac12\int\vec ad\vec l$, where the integration path goes across 
the region where scattering takes place and noise is generated
(e.g. across the barrier in the point contact). 
The sign $\mp$ indicates that the field $\vec a$
is antisymmetric on the upper and lower parts of the Keldysh contour.
Because of that, even though $\vec a$ resembles in many ways 
an ordinary electromagnetic gauge field 
(allowing for gauge transformations, etc.), it has no such meaning.
We emphasize that $\vec a$ is really an auxiliary field describing 
coupling with a {\it virtual} measurement device, such as the spin 
$1/2$ above.

The microscopic formula (\ref{eq:Zkeldysh}), originating from the analysis 
of a coupling with spin $1/2$, is in fact adequate for any ideal 
``passive charge detector'' without 
internal dynamics. We shall use this formula below to obtain counting 
statistics for several physical situations of interest, including
tunneling and mesoscopic transport.

This still leaves some questions about universality and limitations 
of Eq.(\ref{eq:Zkeldysh}). Nazarov and Kindermann\,\cite{nazarov-general} 
considered a more general scheme of charge detection and recovered 
the expression (\ref{eq:Zkeldysh}). Although this is reassuring, 
Ref.\,\cite{nazarov-general} concludes that the detector back action 
is inevitable. Thus it is still
desirable to study more realistic models of noise detection 
that include conversion of microscopic current noise into 
electromagnetic field (photons) as well as an amplifier. Some aspects
of electron-to-photon noise conversion were studied
by Beenakker and Schomerus\,\cite{beenakker-photons}.

It is also of interest to compare the back action effects in different models.
We argue that the above scheme  is likely to describe noise measurement with 
the least back action, 
since coupling to the precessing spin $1/2$ affects only 
the phase of electron forward scattering amplitude, without changing 
scattering probabilities.

\subsection{Statistics of the tunneling current}
\label{sec:tunneling}

The problem of the tunneling current noise provides a simple 
test for the microscopic formula (\ref{eq:Zkeldysh}).
The starting point of our analysis will be the tunneling Hamiltonian
$\hat{\cal H}=\hat{\cal H}_1+\hat{\cal H}_2+\hat V$, 
where $\hat{\cal H}_{1,2}$ describe the 
leads and $\hat V=\hat J_{12}+\hat J_{21}$ is the tunneling operator. 
The specific form of the operators $\hat J_{12}$, $\hat J_{21}$ that 
describe tunneling of a quasiparticle 
between the leads will not be important for the most of our discussion. 
Both the discussion and the results for the tunneling current statistics 
obtained in this section are valid for a generic 
interacting many-body system.

The counting field $\vec a$ in this case can be taken localized on the barrier,
entering the Hamiltonian through the phase factors 
$\exp(\pm i\int \vec ad\vec l)=\exp(\pm i\lambda/2)$
of the operators $\hat J_{12}$, $\hat J_{21}$. The tunneling
operator then is 
  \be\label{Vlambda}
\hat V_{\lambda}= e^{\frac{i}2\lambda(t)}\hat J_{12}(t)
+ e^{-\frac{i}2\lambda(t)}\hat J_{21}(t)\,.
  \ee
Here $\lambda(t)=\pm\lambda$ is antisymmetric on the Keldysh contour $C_{0,t}$.

In what follows we compute $\chi(\lambda)$ and establish a relation with 
the Kubo theorem for the tunneling current\,\cite{mahan}. 
For that, we perform the usual
gauge transformation turning the bias voltage into the tunneling 
operator phase factor as
$\hat J_{12}\rightarrow \hat J_{12}e^{-ieVt}$,
$\hat J_{21}\rightarrow \hat J_{21}e^{ieVt}$. 
Passing to the Keldysh interaction representation, we write
  \be\label{eq:chi-V}
\chi(\lambda) = \Big\langle {\rm T}_{\rm K} 
\exp\lp -i\int_{C_{0,t}}\hat V_{\lambda(t')}(t')dt'\rp\Big\rangle\,.
  \ee
Diagrammatically, the partition function (\ref{eq:chi-V}) is a sum 
of linked cluster diagrams with appropriate combinatorial factors. 
To the lowest order in the tunneling operators $\hat J_{12}$, $\hat J_{21}$
we only need to consider linked clusters of the second order.
This gives 
  \be\label{eq:W-int}
\chi(\lambda) =e^{W(\lambda)},\quad
W(\lambda)=-\frac12   
\int\!\!\int_{C_{0,t}}
\!\Big\langle {\rm T}_{\rm K} \hat V_{\lambda(t')}(t')
\hat V_{\lambda(t'')}(t'')\Big\rangle\,
dt'dt'' .
  \ee
This result is correct for the measurement time $t$ much larger than 
the correlation time in the contacts that determines the characteristic 
time separation $t'-t''$ at which the correlator in (\ref{eq:W-int}) 
decays.

There are several different contributions to the integral in 
(\ref{eq:W-int}), arising from $t'$ and $t''$ taken 
on the forward or backward  
parts of the contour $C_{0,t}$. Evaluating them separately, we obtain
  \bea\label{Wlambda2}
&& W(\lambda)= 
\int_0^t\!\!\int_0^t\!\Big\langle \hat V_{-\lambda}(t')
\hat V_{\lambda}(t'')\Big\rangle\,
dt''dt' 
\\ \nonumber
&& \qquad -\!\int_0^t\!\!\int_0^{t'}\!\!\Big\langle\! \hat V_{\lambda}(t')
\hat V_{\lambda}(t'')\!\Big\rangle\,
dt''\!dt'
-\!\int_0^t\!\!\int_{t'}^t\!\!\Big\langle\! \hat V_{-\lambda}(t')
\hat V_{-\lambda}(t'')\!\Big\rangle\,
dt''\!dt' .
  \eea
We substitute the expression (\ref{Vlambda}) in Eq.(\ref{Wlambda2})
and average by pairing $\hat J_{12}$ with $\hat J_{21}$. This gives
  \be\label{eq:Wlambda}
W(\lambda)= 
(e^{i\lambda}\!-\!1)N_{1\to 2}(t)+(e^{-i\lambda}\!-\!1)N_{2\to 1}(t)
  \ee
with
\be\label{eq:N12}
N_{j\to k}=
\int_0^t\!\!\int_0^t \langle \hat J_{kj}(t')\hat J_{jk}(t'')\rangle\, dt'dt'' 
=g_{jk}t
\ee
%
the mean charge   
transmitted from the contact $j$ to the contact $k$ in a time $t$. 
Exponentiating Eq.(\ref{eq:Wlambda}) gives 
nothing but the bidirectional Poisson distribution $\chi_{2P}(\lambda)$
defined by Eq.(\ref{eq:chi-bidirectional})
with the transition rates given by $\bar n=N_{1\to 2}=g_{12}t$, 
$\bar n'=N_{2\to 1}=g_{21}t$,
respectively.

Eq.(\ref{eq:chi-bidirectional}) yields interesting relations between 
different statistics of the distribution.
The cummulants 
$\langle\!\langle\delta n^k\rangle\!\rangle $, obtained by expanding
$\ln \chi_{2P}(\lambda)$ in $\lambda$, are
  \be\label{eq:qmoments}
m_k=\langle\!\langle\delta n^k\rangle\!\rangle =\,\cases{(g_{12}-g_{21})t,\quad & $k$ odd;\cr
(g_{12}+g_{21})t,\quad & $k$ even.}
  \ee
Setting $k=1,2$ we express $g_{12}\pm g_{21}$ through the time-averaged 
current and the low frequency noise spectral density\footnotemark{}:  
\footnotetext{
  The spectral density of the noise is defined through the 
symmetrized current correlator as
${\cal S}_\omega=
\int\left\{\overline{\delta I(t), \delta I(0)}\right\}_+ e^{i\omega t}dt$. 
At $\omega=0$, one can write ${\cal S}_0$ in terms of the variance 
of charge $q(\tau)=\int_t^{t+\tau} I(t')dt'$ transmitted during a long time $\tau$ as
${\cal S}_0=\frac2{\tau}\overline{\delta q^2(\tau)}$.}
  \be\label{eq:IP}
g_{12}-g_{21}=I/q_0, \quad
g_{12}+g_{21}= {\cal S}_0/2q_0^2 ,
  \ee
with $q_0$ the tunneling charge. 
Of special interest for us will be the third correlator
of the transmitted charge 
  \be
\langle\!\langle\delta q^3\rangle\!\rangle 
\equiv \overline{\lp\delta q-\overline{\delta q}\rp^3} . 
  \ee
For this correlator
Eq. (\ref{eq:qmoments}) gives 
$\langle\!\langle\delta q^3\rangle\!\rangle =C_3t$ with the 
coefficient $C_3$
(the spectral density of the third correlator at $\omega=0$) related to  
current as
  \be\label{eq:q3-I}
C_3\equiv \langle\!\langle\delta q^3\rangle\!\rangle /t =q_0^2 I .
  \ee
We note that the relation (\ref{eq:q3-I}) holds for the distribution 
(\ref{eq:chi-bidirectional}) 
at any ratio $(g_{12}-g_{21})/(g_{12}+g_{21})$ of the mean transmitted 
charge to the variance.

The quantities (\ref{eq:N12})
have several general properties. 
First, by writing the expectation values (\ref{eq:N12})
in a basis of exact microscopic states and using the 
detailed balance relation, we obtain 
\be
N_{1\to2}/N_{2\to1}\equiv g_{12}/g_{21}=\exp(eV/k_{\rm B}T)\,,
\ee
where $V$ is the voltage applied to the contacts. Using 
this result to calculate the ratio of the first and second cummulants, 
Eq.(\ref{eq:IP}), we have $(g_{12}-g_{21})/(g_{12}+g_{21})=\tanh(eV/k_{\rm B}T)$.
This gives the noise-current relation 
\be\label{eq:Schottky-VT}
{\cal S}_0=2 q_0\,\coth(eV/k_{\rm B}T)\, I
\ee
that holds for arbitrary $eV/k_{\rm B}T$.
This relation was pointed out by Sukhorukov and Loss\,\cite{sukhorukov}.

Also, one can establish a relation of the quantities 
(\ref{eq:N12})
with the Kubo theorem. 
We consider the tunneling current operator 
  \be
\hat{\cal I}(t)
=-iq_0\lp \hat J_{12}(t)-\hat J_{21}(t)\rp. 
  \ee
From the Kubo theorem for the tunneling current\,\cite{mahan}, 
the mean integrated current 
$\int_0^t\langle \hat{\cal I}(t')\rangle dt'$
is
  \be\label{Qave}
q_0\int_0^t\!\!\int_0^t \Big\langle \lb \hat J_{21}(t'),\hat J_{12}(t'')\rb \Big\rangle\,dt'dt''
=q_0\lp N_{1\to 2}-N_{2\to 1}\rp\,.
  \ee
By writing $N_{j\to k}=g_{jk}t$, we confirm the first relation (\ref{eq:IP}).
To obtain the second relation (\ref{eq:IP}) we consider the variance 
of the charge transmitted in time $t$, given by 
$\langle\!\langle\delta q^2\rangle\!\rangle =q_0^2\int_0^t\!\!\int_0^t \langle\{ \hat{\cal I}(t'),\hat{\cal I}(t'')\}_+ \rangle\,dt'dt''$.
This integral can be rewritten as
  \be
\int_0^t\!\!\int_0^t\! \Big\langle\! \left\{ \hat J_{12}(t'),\hat J_{21}(t'')\right\}_+\! \Big\rangle\,dt'dt''
= N_{1\to 2}\!+\! N_{2\to 1}\,,
  \ee
which immediately leads to the second relation (\ref{eq:IP}).

We conclude that the tunneling current statistics, described by 
Eq.(\ref{eq:chi-bidirectional}), are simpler than in a generic system. 
The current-noise relation, typically known in a generic system only at
equilibrium (Nyquist) and in the fully out-of-equilibrium (Schottky)
regimes, for the tunneling current is given by Eq.(\ref{eq:Schottky-VT})
at arbitrary $eV/k_{\rm B}T$. 

In contrast, the relation (\ref{eq:q3-I}) obeyed by the third 
correlator (\ref{eq:q3ave}) is completely 
insensitive to the crossover between the Nyquist
and Schottky noise regimes.
The meaning of Eq.(\ref{eq:q3-I}) is similar to that of the Schottky formula
${\cal S}_0=2\langle\!\langle\delta q^2\rangle\!\rangle =2q_0 I$. 
However, the Schottky current-noise relation is valid only when charge flow is 
unidirectional, i.e. at low temperatures $k_{\rm B}T\ll eV$, 
since $g_{12}/ g_{21}=\exp(eV/k_{\rm B}T)$, while Eq.(\ref{eq:q3-I})
holds at any $eV/k_{\rm B}T$. 

In experiment, when the current-noise relation is used to
determine the tunneling quasiparticle charge $q_0$ from the tunneling 
current noise, it is crucial to maintain low temperature $k_{\rm B}T\ll eV$.
The requirement of a cold sample at a relatively high bias voltage
is the origin of a well known difficulty in the noise measurement.
In contrast, the relation (\ref{eq:q3-I}) is not constrained by 
any requirement on sample temperature.

This property of the third moment, if confirmed experimentally, 
may prove to be quite useful for measuring quasiparticle 
charge. In particular, this applies to the situations when the 
$I-V$ characteristic is strongly nonlinear, when it is usually 
difficult to unambiguously interpret the noise {\it versus} current 
dependence as a shot noise effect or as a result 
of thermal noise generated by non-linear conductance. 
This appears to be a completely general problem 
pertinent to any interacting system. Namely, in the systems such as 
Luttinger liquids, the $I-V$ nonlinearities arise at $eV\ge k_{\rm B}T$. 
However, it is exactly this voltage that has to be applied for 
measuring the shot noise in the Schottky regime.

Finally, we note that the universality of the third moment is 
specific for the tunneling problem. In other situations, such as a point 
contact or a mesoscopic system, the third moment is 
temperature-dependent\,\cite{levitov01'2,gutman02,nagaev02}.

\subsection{A relation to the theory of photodetection}
\label{sec:photo-detection}

The statistics of tunneling particles was not specified in the 
above discussion, since everything said so far is good
for both bosons and fermions. To illustrate this, here
we discuss the relation of the present approach to the theory of photon
counting\,\cite{Mandel,Glauber}.
A system of photons
interacting with
atoms in a photon
detector can be accounted for by a Hamiltonian of the form 
$\H=\H_{\rm p}+\H_{\rm a}+V$, where $\H_{\rm p}$ describes
free electromagnetic field, $\H_{\rm a}$ is the Hamiltonian 
of atoms in the detector, and 
\be\label{eq:Vphotons}
\hat V=\sum_{j,k} \lp u_{j,k} e^{\frac{i}2\lambda} b_j^\dagger a_k +
u_{j,k}^\ast e^{-\frac{i}2\lambda} a_k^\dagger b_j \rp
\ee
describes the interaction of photons with the atoms, i.e.
the process of photon absorption and atom excitation. Here
$a_k$ are the canonical Bose operators of photon modes, labeled by $k$,
and $b_j$ are the operators describing excitation of the atoms. 
Since the operator $\hat V$ transfers excitations between the field 
and the atom systems, it can be interpreted as a ``tunneling operator.'' 
(The only difference is
in the unidirectional character of the ``current'' induced by  
$\hat V$, since photons can be only absorbed in the detector 
but not created.)
This analogy allows one to use the formalism of Sec.\ref{sec:micro-spin1/2}
to study photon counting, and for that purpose we
added a counting field in (\ref{eq:Vphotons})
(compare to Eq.(\ref{Vlambda})).

Given all that, the generating function for photons has 
the form (\ref{eq:chi-V}) which we rewrite to show an explicit dependence 
on the measurement time:
\be\label{eq:chi(t)}
\chi_t(\lambda)=\Big\langle \U^{-1}_{-\lambda}(t)\,\U_{\lambda}(t)\Big\rangle
,\qquad
\U_{\lambda}(t) =
{\rm Texp\/}\lp -i\int_0^t\hat V_{\lambda}(t')dt'\rp
\,,
\ee
The task of evaluating the partition function (\ref{eq:chi(t)}) is simplified 
by the weakness of the photon-atom coupling. (Each atom 
is excited during the counting time $t$ with a very small probability.) 
The expression (\ref{eq:chi(t)}) can thus be evaluated by taking into account 
the interaction of a photon with each of the atoms only to the lowest order.
This is also similar to the tunneling problem.

However, at this stage the similarity with tunneling ends, 
since photon coherence time can be much longer than the measurement time $t$. 
The method of Sec.\ref{sec:tunneling}, based on 
the linked cluster expansion of $\ln\chi$, should be modified 
to account for the long coherence times. Another complication is that
the photon density matrix is not specified, since we are not 
limiting the discussion to thermal photon sources. 

We handle the partition function (\ref{eq:chi(t)}) by averaging over
atoms, while keeping the photon variables free. As explained above, 
only pairwise averages of atoms' operators are needed. We write them as 
$\langle b^\dagger_j(t) b_{j'}(t') \rangle = 0$, 
$\langle b_j(t) b_{j'}^\dagger (t') \rangle = \tau_j\delta(t-t')\delta_{jj'}$,
where $\tau_j$ is a constant of the order of the excitation 
time of an atom, and the $\delta$-function is actually a function of the 
width $\simeq\tau_j$. (Typically, $\tau_j$ is a very short, microscopic time.) 

Turning to the calculation, let us consider the difference 
$\chi_{t+\Delta}(\lambda)-\chi_t(\lambda)$, with the time increment
$\Delta$ large compared to  $\tau_j$, but much smaller than the 
characteristic photon coherence time. Expanding 
$\U_{\lambda}(t+\Delta)$
to the second order in $\Delta$, we write it as
\be
\lp 1-i \int_t^{t+\Delta}\hat V_{\lambda}(t')dt'
-\frac12 \int_t^{t+\Delta}\!\!\!\int_t^{t'}\hat V_{\lambda}(t')\hat V_{\lambda}(t'')dt'dt''
\rp\U_{\lambda}(t)\,.
\ee
Substituting this in Eq.(\ref{eq:chi(t)}) along with a similar expression for 
$\U^{-1}_{-\lambda}(t+\Delta)$, and averaging over the atoms as described 
above, we obtain
\be\label{eq:chi-dot}
\partial_t\chi_t(\lambda)=
(\chi_{t+\Delta}(\lambda)-\chi_t(\lambda))/\Delta
=\sum_k\eta_k(e^{i\lambda}-1)\,
\Big\langle \U^{-1}_{-\lambda}(t) a^\dagger_k a_k\,\U_{\lambda}(t)\Big\rangle
\ee
with $\eta_k=\sum_j \tau_j|u_{j,k}|^2$ the detector
efficiency parameters. The solution of Eq.(\ref{eq:chi-dot})
has the form well known in optics\,\cite{Mandel,Glauber}\,:
\be\label{eq:chi-photons-prod}
\chi_t(\lambda)=\prod_k \chi_t^{(k)}(\lambda)
,\quad
\chi_t^{(k)}(\lambda)=\Big\langle :\exp\lp\eta_kt(e^{i\lambda}-1)a^\dagger_k a_k\rp: \Big\rangle_k
\,,
\ee
where $:...:$ is the normal ordering symbol and $\langle...\rangle_k$ 
is the average over photon density matrix. The product rule in
(\ref{eq:chi-photons-prod}) indicates that the counting distributions 
for different electromagnetic modes are statistically independent. 
The physical meaning of
normal ordering
is that each photon, after having been detected, 
is absorbed in the detector and destroyed.

From Eq.(\ref{eq:chi-photons-prod}), the counting probability
of $m$ photons in one mode is
   \be\label{A1}
p_m^{(k)}={(\eta_k t)^m \over m!}\langle\ :(a_k^\dagger a_k)^m e^{-\eta_k t a_k^\dagger a_k}:\ \rangle_k \,.
   \ee
Eqs.(\ref{eq:chi-photons-prod}),(\ref{A1}) are central  
to the theory of photon counting\,\cite{Mandel}. 
Particularly interesting is the case of a
coherent photon state $|{\rm z}\rangle$, $a|{\rm z}\rangle=z\, |{\rm z}
\rangle$, with a complex $z$, corresponding to the
radiation field of an ideal laser. In this case Eq.(\ref{A1}) yields 
the Poisson distribution
$p_m=e^{-Jt}(Jt)^m/m!$,
$J=\eta|z|^2$,
which describes the so-called minimally bunched light sources.

\section{Counting statistics of mesoscopic transport}

Here we consider the problem of counting statistics in a mesoscopic transport.
From now on we adopt the noninteracting particle approximation
and use the scattering approach\,\cite{beenakkerRMP}, 
in which the system is characterized by a single particle scattering matrix. 
Depending on the nature of the problem, 
the matrix can be stationary or time-dependent. Even for noninteracting 
particles the problem of counting statistics remains nontrivial due 
to correlations between different particles arising from Fermi statistics.

Counting statistics can be analyzed using the microscopic formula
(\ref{eq:Zkeldysh}). However, there is a more efficient way of handling 
the noninteracting problem. One can obtain a formula for the generating 
function $\chi(\lambda)$ in terms of a functional determinant that 
involves the scattering matrix and the density matrix of reservoirs.  
Then for each particular problem one has to analyze and evaluate an 
appropriate determinant. Although functional determinants 
can be nontrivial to deal with, this approach is still much simpler 
than the one based directly on Eq.(\ref{eq:Zkeldysh}).

\subsection{Statistics of the DC transport}
\label{sec:DC}

He we discuss the problem of time-independent scattering. 
We consider a conductor with $m$ scattering channels describing 
states within one or several current leads. The scattering is elastic 
and will be characterized by a $m\times m$ matrix $S$. Although in 
applications so far the $2\times2$ 
matrices (i.e. the problems with two scattering channels)
have been more common than larger matrices, 
the general determinant structure of $\chi(\lambda)$ will 
be revealed only for matrices of arbitrary size $m$.

For elastic scattering
one can obtain $\chi(\lambda)$ from a quasiclassical argument.
In this case, particles with different energies contribute 
to counting statistics independently, and thus one can ``symbolically'' write
\be\label{eq:chi-prod}
\chi(\lambda)=\prod_\epsilon \chi_\epsilon(\lambda)
\,,\quad {\rm i.e.}\quad
\chi(\lambda)=\exp\lp t\int \ln\chi_\epsilon(\lambda)
\frac{d\epsilon}{2\pi\hbar}\rp
 ,
\ee
where $\chi_\epsilon(\lambda)$ is the contribution of particles with energy 
$\epsilon$. The factor $2\pi\hbar$ is written based
on the quasiclassical phase space volume normalization, 
$d{\cal V}=d\epsilon dt/2\pi\hbar$.
The quantity $\chi_\epsilon(\lambda)$ depends on the 
scattering matrix $S$ and on the energy distribution $n_i(\epsilon)$ 
in the channels.

To obtain $\chi_\epsilon(\lambda)$ we introduce a vector of counting fields
$\lambda_j$, $j=1,...,m$, one for each channel, and consider all possible
multi-particle scattering processes at fixed energy. The processes
can involve any number $k\le m$ of particles each coming out of one of 
the $m$ channels and being scattered into another channel. Since the 
particles are indistinguishable fermions, no two particles can share 
an incoming or outgoing channel. One can then write $\chi_\epsilon(\lambda)$
as a sum over all different multiparticle scattering processes:
\be\label{eq:chi-mm}
\chi_\epsilon(\lambda)=\sum\limits_{i_1,...,i_k,j_1,...,j_k}
e^{\frac{i}2(\lambda_{i_1}+...+\lambda_{i_k}-\lambda_{j_1}-...-\lambda_{j_k})}
P_{i_1,...,i_k\,|\,j_1,...,j_k}
\, ,
\ee
where the rate of $k$ particles transition from channels $i_1,...,i_k$
into channels $j_1,...,j_k$ is given by
\be
P_{i_1,...,i_k\,|\,j_1,...,j_k}=\Big|S_{i_1,...,i_k}^{j_1,...,j_k}\Big|^2
\prod_{i\ne i_\alpha}(1-n_i(\epsilon)) \prod_{i= i_\alpha}n_i(\epsilon)
\, .
\ee
Here $S_{i_1,...,i_k}^{j_1,...,j_k}$ 
is an antisymmetrized product of $k$ single particle amplitudes,
which is nothing but the minor of the matrix $S$ with rows $j_1,...,j_k$
and columns $i_1,...,i_k$. The product of $n_i$ and $1-n_i$
gives the probability to have $k$ particles come out of the channels 
$i_1,...,i_k$.

An important insight in the structure of the expression (\ref{eq:chi-mm}) 
can be 
obtained by noting that it has a form of a determinant:
\be\label{eq:chi-det-mm}
\chi_\epsilon(\lambda)={\rm det\,}\lp \hat 1 -\hat n_\epsilon 
+ \hat n_\epsilon S^{-1}_{-\lambda}S_{\lambda}\rp
,\qquad
(S_{\lambda})_{ij}=e^{\frac{i}4(\lambda_i-\lambda_j)}S_{ij} .
\ee
Here $\hat n_\epsilon$ is a diagonal $m\times m$ matrix of channel occupancy 
at the energy $\epsilon$ and 
the counting field 
$\lambda_j$ enters in the phase factors of the matrix $S_{\lambda}$. To demonstrate that the expressions 
(\ref{eq:chi-mm}) and (\ref{eq:chi-det-mm}) are identical
one has to expand the determinant (\ref{eq:chi-det-mm}) 
and go through a bit of matrix algebra. The formula (\ref{eq:chi-det-mm})
is particularly useful because, as we shall see below, it can be generalized 
to time-dependent problems.

Let us now focus on the simplest case $m=2$ which describes transport in 
a point contact, with the two channels corresponding to current
leads. The $2\times2$ scattering matrix $S$ contains reflection and 
transmission amplitudes. In this case, since there are only six terms in
Eq.(\ref{eq:chi-mm}), the determinant formula (\ref{eq:chi-det-mm}) 
is not necessary. From Eq.(\ref{eq:chi-mm}) we obtain
\bea\label{eq:chi-22}
&&\chi_\epsilon(\lambda) = (1-n_1)(1-n_2)
+(|S_{11}|^2+e^{\frac{i}2(\lambda_2-\lambda_1)}|S_{21}|^2)n_1(1-n_2) 
\nonumber \\
&& \phantom{\chi_\epsilon(\lambda)}
+(|S_{22}|^2+e^{\frac{i}2(\lambda_1-\lambda_2)}|S_{12}|^2)n_2(1-n_1)
+|{\rm det\,}S|^2 n_1 n_2
\,,
\eea
where the energy dependence of $n_j(\epsilon)$ is suppressed.
By using the unitarity relations
$|S_{1i}|^2+|S_{2i}|^2=1$, $|{\rm det\,}S|=1$, Eq.(\ref{eq:chi-22}) can be simplified:
\be
\chi_\epsilon(\lambda) = 
1+p(e^{i\lambda}-1)n_1(1-n_2)+p(e^{-i\lambda}-1)n_2(1-n_1)
\,.
\ee
Here $p=|S_{21}|^2=|S_{12}|^2$ is the transmission coefficient and 
$\lambda=\lambda_2-\lambda_1$. 
(We denote the transmission probability by $p$ instead 
of a more traditional $t$ to avoid confusion with the measurement time.)

To obtain the full counting statistics integrated over all energies,
one has to specify the energy distribution in the leads and 
use Eq.(\ref{eq:chi-prod}).
We consider a barrier with energy-independent transmission and the 
leads at temperature $T$ biased by voltage $V$. 
Then $n_{1,2}=n_F(\epsilon\mp eV/2)$ with $n_F$ the Fermi function.

At $T=0$, since $n_F(\epsilon)$ takes values $0$ and $1$,
for $V>0$ we have
\be
\chi_\epsilon(\lambda) = 
\cases{e^{i\lambda}p+1-p\,, & $|\epsilon|<\frac12 eV$;\cr
1\,, & $|\epsilon|>\frac12 eV$. }
\ee
After doing the integral in Eq.(\ref{eq:chi-prod}) we obtain the binomial
distribution (\ref{eq:chi-binomial}), $\chi_N(\lambda)=(e^{i\lambda}p+1-p)^{N(t)}$, 
with the number of attempts 
$N(t)=eVt/2\pi\hbar$.

This means that, in agreement with intuition, 
in the energy window $eV$ the transport is just 
the single particle transmission and reflection, while the states 
with energies in the Fermi sea, populated in
both reservoirs, are noiseless. (At $V<0$ the result is similar, 
with $e^{i\lambda}$ replaced by $e^{-i\lambda}$, 
which corresponds to the DC current sign reversal.)

We note that the noninteger number of attempts $N(t)=eVt/2\pi\hbar$ 
is an artifact of a quasiclassical calculation. In a more careful analysis
the number of attempts is characterized by a narrow distribution $P_N$
peaked at $\overline{N}=N(t)$, and the generating function is a weighted sum
$\sum_N P_N \chi_N(\lambda)$. Since the peak width is a sublinear function 
of the measurement time $t$ (in fact, $\overline{\delta N^2}\propto \ln t$), 
the statistics to the leading order in $t$
are correctly described by the binomial distribution.

One can also consider the problem at 
arbitrary $k_{\rm B}T/eV$\,\cite{levitov94}.
The integral in Eq.(\ref{eq:chi-prod}), although less trivial, can 
be carried out, giving
  \begin{equation} \label{A24}
{\chi}({\lambda})=
\exp\left(- u_+u_- N_T\right)\, ,\quad N_T=t\,k_{\rm B}T/2\pi\hbar,
  \end{equation}
where
\be
u_\pm=v\pm\cosh^{-1}(p\cosh(v+i\lambda)+(1-p)\cosh v)
,\quad
v=eV/2k_{\rm B}T
\,.
\ee
At low temperature $ k_{\rm B}T\ll eV$, Eq.(\ref{A24}) reproduces 
the binomial statistics.
At low voltage $eV\ll k_{\rm B}T$ (or high temperature) 
Eq.(\ref{A24}) gives the
counting statistics of the equilibrium Nyquist noise:
\be
\label{A23}
\chi(\lambda)= e^{-\lambda_*^2 N_T},
\qquad 
\sin(\lambda_*/2)=p^{1/2}\sin(\lambda/2)\,.
\ee
Remarkably, even at equilibrium
the noise is non-gaussian, except for special case of full transmission, 
$p=1$, $\lambda_*=\lambda$, when it becomes
gaussian. 

\subsection{Statistics of time-dependent scattering}

The time-dependent scattering problem describes photon-assisted transport.
There are two groups of practically interesting problems: 
the AC-driven systems with static scattering potential, such as 
tunneling barriers or point contacts in the presence of a microwave 
field\,\cite{schoelkopf98,glattli02}, 
and the electron pumps with time-dependent scattering 
potential controlled externally, e.g. by gate voltages\,\cite{Switkes99}.

Typically, the time of individual particle transit through 
the scattering region is much shorter than the period at which the system 
is driven. This situation is described, in the instantaneous 
scattering approximation, by a time-dependent scattering matrix 
$S(t)$ that characterizes single particle scattering at time $t$.
The question of interest is how Fermi statistics of many-body 
scattering states affects the counting statistics. 

One can construct a theory of counting statistics of time-dependent 
scattering\,\cite{ivanov93} by generalizing the results of Sec.\ref{sec:DC} for 
the statistics of a generic time-independent scattering. In particular, 
as we discuss below, the determinant formula (\ref{eq:chi-det-mm}), 
along with Eq.(\ref{eq:chi-prod}), 
allows a straightforward extension to time-dependent problems.
In that, the generating function $\chi(\lambda)$ acquires a form of a
functional determinant.

Let us consider a scattering matrix $S(t)$ varying periodically in time 
with frequency $\Omega$.
The analysis is most simple in the frequency representation\,\cite{ivanov93},
in which the scattering operator $S$ has off-diagonal matrix elements 
$S_{\omega',\omega}$ with discrete frequency change 
$\omega'-\omega=n\Omega$.
In this approach the energy axis is divided into intervals 
$n\Omega<\omega<(n+1)\Omega$ and each such interval 
is treated as a separate conduction channel. 
In doing so it is convenient (and in some cases necessary) 
to assign a separate counting field $\lambda_n$ to each frequency channel, 
giving the counting field a frequency 
channel index in addition to the conduction channel dependence 
displayed in Eq.(\ref{eq:chi-det-mm}). 

Since the scattering operator conserves energy modulo multiple 
of $\hbar \Omega$, the scattering can be viewed as {\it elastic} 
in the extended channel representation, which allows to employ 
the method of Sec.\ref{sec:DC}.
We note also that the form of the determinant in Eq.(\ref{eq:chi-det-mm}) 
is not particularly sensitive to the size of the scattering matrix. 
Thus one can use it even when the number of channels is infinite, 
provided that the determinant remains well defined.
This procedure brings (\ref{eq:chi-det-mm}) to the form of a determinant 
of an infinite size matrix.
Determinant regularization can be accomplished by
truncating this matrix 
at very high and low frequencies, thereby eliminating 
empty states and the states deep in the Fermi sea which do not contribute to 
noise. 

Finally, we note that the product rule (\ref{eq:chi-prod}) 
for $\chi(\lambda)$ 
is consistent with the determinant structure, since scattering 
processes at the energies different modulo $n\Omega$ are decoupled.
This allows to keep the answer for $\chi(\lambda)$ in the form of
the determinant (\ref{eq:chi-det-mm}), where now the scattering operator 
$S$ is considered in the entire frequency domain, rather than at  
the discrete frequencies $\omega+n\Omega$. The resulting functional 
determinant has a simple form in the time representation:
\be
\label{eq:chi-det}
  \chi(\lambda)=\ {\rm det}\lp \hat 1 +
\hat n(t,t') \lp {\hat T}_\lambda(t)-\hat 1\rp \rp
,\quad
{\hat T}_\lambda(t)=S^\dagger_{-\lambda}(t)S_{\lambda}(t),
\ee
where $(S_{\lambda})_{jj'}=e^{\frac{i}4(\lambda_j-\lambda_{j'})}S_{jj'}$
as above, and $\hat n$ is the density matrix of reservoirs. 
The operator $\hat n$, diagonal in the channel index,
is given by 
\be
n_{jj'}(t,t')=
\delta_{jj'}\int n_j(\hbar\omega)\, e^{i\omega(t'-t)}\,d\omega/2\pi\,.
\ee
In general $\hat n(t,t')$
depends on the energy distribution parameters, such as temperature
and chemical potential.
In equilibrium, 
by taking Fourier transform of the Fermi function $n_F(\epsilon-\mu)$, 
one obtains
\be\label{eq:n-VT}
n_j(t,t')=
\frac{e^{-i\mu_j(t-t')/\hbar}}{2\beta\sinh(\pi(t-t'+i\delta)/\beta)}
,\quad
\beta=\hbar/k_{\rm B}T\,.
\ee
Finite bias voltage $V$ s described by $\mu_1-\mu_2=eV$. 
At $T=0$ and $V=0$, Eq.(\ref{eq:n-VT}) gives
$n(t,t')=i/(2\pi(t-t'+i\delta))$.
The result (\ref{eq:chi-det}) holds for an arbitrary (even nonequilibrium)
energy distribution in reservoirs. 

The functional determinant of an infinite matrix
(\ref{eq:chi-det}) should be handled carefully.
One can show that, in a mathematical sense, 
the quantity (\ref{eq:chi-det}) is well defined. 
For the states with energies deep in the Fermi sea, $\hat n=1$ and, since
${\rm det}\lp {\hat T}_\lambda(t)\rp=1$ due to unitarity of $S$, 
these states do not contribute to the determinant (\ref{eq:chi-det}).
Similarly, since $\hat n=0$ for the states with very high energy,
these states also do not affect the determinant. 
Effectively, the determinant is controlled by a group of states 
near the Fermi level, in agreement with intuition about transport 
in a driven system. The absence of 
ultraviolet divergences allows one to go freely between 
different representations, e.g. to switch from the frequency domain 
to the time domain, which facilitates calculations\,\cite{levitov96,ivanov97}.  

The above derivation of the formula (\ref{eq:chi-det}) based on
a generalization of the result (\ref{eq:chi-det-mm}) for time-independent 
scattering might seem not entirely rigorous. A more mathematically sound 
derivation that starts directly from the microscopic expression 
(\ref{eq:Zkeldysh}) was proposed recently by Klich\,\cite{klich02}.

\subsection{Case studies}
\label{sec:cases}

Here we briefly review the time-dependent scattering problems for which 
the counting statistics have been studied. From several examples for which 
$\chi(\lambda)$ has been obtained it appears that the problem does not allow
a general solution. Instead, the problem can be handled only for 
suitably chosen form of the time dependence $S(t)$.

In Ref.\,\cite{ivanov93} a two channel problem was considered with
$S(t)$ of the form
  \be
  \label{matrix-93}
S(\tau)\equiv\lp\matrix{r & t' \cr t & r' }\rp
=\lp\matrix{B+be^{-i\Omega \tau} & \bar A+\bar ae^{i\Omega \tau} \cr 
A+ae^{-i\Omega \tau} & -\bar B-\bar be^{i\Omega \tau} }\rp,
  \ee
which is unitary for any $t$ provided 
$|A|^2+|a|^2+|B|^2+|b|^2=1$, $A\bar a+B\bar b=0$. 
The problem was solved by using the extended channel representation 
in the frequency domain, in which each frequency interval 
$n\Omega<\omega<(n+1)\Omega$ is treated as a separate scattering channel, as 
discussed above. 

For the reservoirs at zero temperature and without bias voltage  
the charge distribution for $m$ pumping cycles is described by
  \be
  \label{chi-matrix-93}
\chi(\lambda)=\lp 1+p_1 (e^{i\lambda}-1)+p_2 (e^{-i\lambda}-1) \rp^m 
  \ee
with $p_1=|a|^4/(|a|^2+|b|^2)$ and $p_2=|b|^4/(|a|^2+|b|^2)$. 
This result means that at each pumping cycle one electron is pumped 
in one direction with probability $p_1$, or in the opposite direction
with probability $p_2$, or no charge is pumped with probability $1-p_1-p_2$. 
The multiplicative dependence of $\chi(\lambda)$ on the number 
of pumping cycles $m$ indicates that
the outcomes of different cycles are statistically independent. 
One can thus view (\ref{chi-matrix-93}) 
as a generalization of the binomial distribution (\ref{eq:chi-binomial}).

The problem (\ref{matrix-93}) was also studied in Ref.\,\cite{ivanov93}
at a finite bias voltage, when the counting distribution is not as simple as 
(\ref{chi-matrix-93}). To describe the result, for a given bias voltage $V$
we find an integer $n$ such that $nf< \frac{e}{h}V \le (n+1)f$,
where $f=\Omega/2\pi$ is the cyclic frequency in (\ref{matrix-93}). 
Then for a long measurement time $t\gg\Omega^{-1}$ the counting distribution is
\be\label{eq:chi-n-n+1}
\chi(\lambda)=\chi_n^{N_>}(\lambda)\cdot\chi_{n+1}^{N_<}(\lambda)
\,,
\ee
where
$N_<=\lp {\textstyle \frac{e}{h}}V \!-\! nf\rp t$,
$N_>=\lp (n+1)f \!-\! {\textstyle \frac{e}{h}}V\rp t$,
and the functions $\chi_n(\lambda)$ are finite degree 
polynomials in $e^{\pm i\lambda}$. 
The form of $\chi_n(\lambda)$ depends  
on $A$, $B$, $a$, and $b$ (we refer to Ref.\,\cite{ivanov93} for details). 

The product rule (\ref{eq:chi-n-n+1}) means that
the cummulants of the distribution $\chi(\lambda)$ depend on
$V$ in a piecewise linear way, $m_k(V)=N_>m_k^{(n)}+N_<m_k^{(n+1)}$,
with cusp-like singularities at $eV=nhf\equiv n\hbar\Omega$. 
These singularities are generic for the noise in photo-assisted
phase-coherent transport\,\cite{LevLes94,schoelkopf98,glattli02}.

Another time-dependent problem for which solution can be obtained 
in a closed form is mesoscopic transport in the presence of an AC 
voltage\,\cite{levitov96}. 
The scatterer in this case is a time-independent
$2\times2$ matrix, while the voltage $V(t)$
enters in the phase factors of the density matrix in (\ref{eq:chi-det}):
\be
n_{1,2}(t,t')=e^{\pm\frac{i}2(\varphi(t')-\varphi(t))}n^{(0)}_{1,2}(t,t'),
\quad 
\dot\varphi(t)={\textstyle\frac{e}{\hbar}}V(t)
\ee
(compare this with the formula (\ref{eq:n-VT}) for constant bias voltage).

The counting distribution (\ref{eq:chi-det}) for a family of such problems has
been studied in Ref.\,\cite{ivanov97}. It was noted earlier\,\cite{hwlee95'1} that noise
is minimized at fixed transmitted charge for 
a special form of time-dependent voltage:
\be\label{eq:L-pulses}
V(t)=\frac{h}{e} \sum_{k=1,...,m} 
\frac{2\tau_k}{(t-t_k)^2+\tau_k^2}\,.
\ee
Each of the Lorentzian voltage pulses (\ref{eq:L-pulses}) corresponds 
to a $2\pi$ phase change in $\varphi(t)$.
Interestingly, the noise-minimizing pulses (\ref{eq:L-pulses}) 
have large degeneracy:
they produce noise which is insensitive to the pulses' widths $\tau_k$ and 
peak positions $t_k$. This calls for an interpretation
of the pulses (\ref{eq:L-pulses}) as independent attempts to transmit charge.
Not surprisingly, the counting statistics for such pulses
was found to be binomial:
\be
\chi(\lambda)=(1+t(e^{i\lambda}-1))^m
\ee
with $t$ the transmission constant. The lowest possible noise 
for a current pumped by voltage pulses is thus equal to that of 
a DC current with the same transmitted charge. 

The method of Ref.\,\cite{ivanov97} also allows to find the distribution
for an arbitrary sum of the pulses (\ref{eq:L-pulses}) with alternating signs. 
For example, two opposite pulses
\be
V(t)=\frac{h}{e} \lp 
\frac{2\tau_1}{(t-t_1)^2+\tau_1^2}
-
\frac{2\tau_2}{(t-t_2)^2+\tau_2^2}
\rp
\ee
give rise to the counting distribution
\be
\chi(\lambda)=1-2F+F(e^{i\lambda}+e^{-i\lambda})
\,,\quad
F=t(1-t)\Big|\frac{z_1^\ast-z_2}{z_1-z_2}\Big|^2
\ee
with $z_{1,2}=t_{1,2}+i\tau_{1,2}$. 
The quantity $|...|^2$ is a measure of pulses' overlap in time,
varying between $0$ for a full overlap and $1$ for no overlap. 
For nonoverlapping pulses, $\chi(\lambda)$ factors into
$(t\,e^{i\lambda}+1-t)(t\,e^{-i\lambda}+1-t)$, in agreement with the 
interpretation of
a binomial distribution for independent attempts.

\subsection{Mesoscopic pumping}
\label{sec:mesopump}

A DC current in a mesoscopic system, such as an open quantum dot,
can be induced by pumping, i.e. by modulating its area, shape, or other
parameters\,\cite{Spivak95,Brouwer98,Altshuler'99}. After pumping was 
demonstrated experimentally\,\cite{Switkes99}, it came into the focus of
mesoscopic literature (for references see\,\cite{Altshuler'99,polianski02}). 
In particular, 
Brouwer made an interesting observation
that the time-averaged
pumped current is a purely geometric property of the path in the
scattering matrix parameter space, insensitive to path parameterization.

Transport through a mesoscopic system is described\,\cite{beenakkerRMP} by
a scattering matrix $S$ which depends on externally driven
parameters and varies cyclically with time. 
The matrix $S(t)$ defines a path in the space of all scattering matrices. 
For a system with $m$ scattering channels, the matrix space 
is the group $U(m)=SU(m)\times U(1)$. 
In an experiment one can, in principle, realize any path 
in the space of scattering matrices. 

Counting distribution for a parametrically driven open system was discussed
by Andreev and Kamenev who adapted the results\,\cite{ivanov93} 

obtained for specific pumping cycles [Eqs.(\ref{matrix-93}),(\ref{chi-matrix-93})]. However,
since the relation between the path in the scattering matrix space and the 
external pumping parameters is generally unknown, only the results 
valid for generic paths are of interest in this problem.

Here we consider the weak pumping regime, when the path $S(t)$ is 
a sufficiently small, but otherwise arbitrary loop, 
and show that in this case the counting distribution
is universal\,\cite{levitov01'1}, taking the form of 
bidirectional Poisson distribution (\ref{eq:chi-bidirectional}). 
From that, we obtain the dependence of the noise
on the amplitude and relative phase of the voltages driving the pump.

Before turning to the calculation, we discuss
general dependence 
of counting statistics on the path in matrix space. 
Different paths $S(t)$, in principle, give rise to different 
current and noise. However, there is a remarkable property 
of invariance with respect to group shifts. Any two paths, 
  \be
  \label{invariance}
S(t)\quad {\rm and}\quad S'(t)=S(t)S_0, 
  \ee
where $S_0$ is a time-independent matrix in $U(m)$, give rise to 
the same counting statistics at zero temperature. 
We note that only the right shifts of the form (\ref{invariance}) leave 
counting statistics invariant, whereas the left shifts generally change it. 
One can explain the result (\ref{invariance}) qualitatively as follows.
The change of scattering matrix, $S(t)\to S'(t)=S(t)S_0$, is equivalent to
replacing states in the {\it incoming} scattering channels by their 
superpositions $\psi^\alpha=S_{0\beta}^\alpha \psi^\beta$. 
At zero temperature, however, Fermi reservoirs are 
noiseless and also such are any their superpositions. Correlation between
superposition states of noiseless reservoirs is negligible, while
current fluctuations 
arise only due to the time-dependent scattering. Therefore, noise statistics
remain unchanged. A simple formal proof of the result (\ref{invariance})
is given below. 

For a weak pumping field it is sufficient to evaluate (\ref{eq:chi-det}) 
in the time domain 
by expanding $\ln {\rm det}(...)$ in powers of $\delta S$ and keeping
non-vanishing terms of lowest order. In doing so, however, we preserve
full functional dependence on $\lambda$ which gives
all moments of counting statistics. We write
$S(t)=e^{A(t)}S^{(0)}$ with antihermitian $A(t)$ 
representing small perturbation, ${\rm tr} A^\dagger A\ll1$.
Here $S^{(0)}$ is scattering matrix of the system in the absence of 
pumping. Substituting this into (\ref{eq:chi-det}) one obtains
  \be
\label{T-lambda}
{\hat T}_\lambda(t)\equiv
{\hat T}_\lambda^{(0)}+\delta T_\lambda(t)=
S^{(0)\dagger}_{-\lambda} e^{-A_{-\lambda}(t)}e^{A_{\lambda}(t)}S^{(0)}_{\lambda}
  \ee
with ${\hat T}_\lambda^{(0)}=S^{(0)\dagger}_{-\lambda} S^{(0)}_{\lambda}$
and
$A_{\lambda}(t)=
e^{i\frac{\lambda}4\sigma_3}A(t)e^{-i\frac{\lambda}4\sigma_3}$
(here $\sigma_3$ equals $+1$ for the left and $-1$ for the right channel). 
Now, we expand (\ref{eq:chi-det}):
  \be
\label{ln-chi}
\ln \chi(\lambda)=\ln {\rm det} Q_{0}+ 
{\rm tr} R
-\frac12 {\rm tr} R^2
+\frac13 {\rm tr} R^3
-... \,,
  \ee
where $Q_{0}=1+\hat n({\hat T}_\lambda^{(0)}-1)$
and $R=Q_{0}^{-1}\hat n\delta T_{\lambda}$. 
At zero temperature, from $\hat n^2=\hat n$ it follows that
${\rm det} Q_{0}=1$ 
and $R=S^{(0)\!\ -1}_{\lambda}
\hat n \lp e^{-A_{-\lambda}(t)}e^{A_{\lambda}(t)}-1\rp S^{(0)}_{\lambda}$. 
Therefore,
  \be
\label{ln-M}
\ln \chi(\lambda)= {\rm tr}\ \hat n\hat M
-\frac12 {\rm tr} (\hat n\hat M)^2
+\frac13 {\rm tr} (\hat n\hat M)^3
-... \,,
  \ee
where $\hat M=e^{-A_{-\lambda}(t)}e^{A_{\lambda}(t)}-1$. 
Note that at this stage there is no dependence left on
the constant matrix $S^{(0)}$, which proves the invariance 
under the group shifts (\ref{invariance}). 

We need to expand (\ref{ln-M})
in powers of the pumping field, which amounts to taking 
the lowest order terms of the expansion in powers of
the matrix $A(t)$. One can check that the two ${\cal O}(A)$ terms 
arising 
from the first term in (\ref{ln-M}) vanish. 
The ${\cal O}(A^2)$ terms 
arise from the first and 
second term in (\ref{ln-M}) and have the form
   \be
\label{AB}
\ln \chi=  
\frac12 {\rm tr}\lp \hat n \lp
A_{-\lambda}^2 +A_{\lambda}^2 -\! 2 A_{-\lambda}A_{\lambda}\rp\rp
\! -\frac12 {\rm tr} (\hat n B_{\lambda})^2
  \ee
with $B_{\lambda}(t)=A_{\lambda}(t)-A_{-\lambda}(t)$. 
At zero temperature, by using 
$\hat n^2=\hat n$, one can bring (\ref{AB}) to the form
   \be
\label{ln-AB}
\frac12\!\ {\rm tr}\!\ \lp\hat n 
\lb A_{\lambda},A_{-\lambda}\rb\rp
+\frac12 \lp {\rm tr} \lp \hat n^2 B_{\lambda}^2 \rp
- {\rm tr} (\hat n B_{\lambda})^2 \rp .
  \ee
The first term of (\ref{ln-AB}) has to be regularized in 
the Schwinger anomaly fashion, by splitting points, 
$t',t''=t\pm \eta/2$, which gives
   \be
   \label{anomaly}
\frac12 \oint n(t',t'')
{\rm tr} \lp A_{-\lambda}(t'') A_{\lambda}(t')
-A_{\lambda}(t'') A_{-\lambda}(t') \rp 
dt \,.
  \ee
Averaging over small $\eta$ can be achieved either by
inserting in (\ref{anomaly}) additional integrals over 
$t'$, $t''$, or simply by replacing 
$A_{\lambda}(t)\to \frac12\lp A_{\lambda}(t)+A_{\lambda}(t')\rp$, etc.
After taking the limit $\eta\to 0$,
Eq.(\ref{anomaly}) becomes
   \be
\label{int1}
\frac{i}{8\pi} \oint {\rm tr}\lp 
A_{-\lambda}\partial_t A_{\lambda}
-
A_{\lambda}\partial_t A_{-\lambda}
\rp dt\,.
   \ee
The second term of (\ref{ln-AB}) can be written as
   \be
\label{int2}
\frac1{4(2\pi)^2} \oint\!\!\oint 
\frac{{\rm tr} \lp B_{\lambda}(t)-B_{\lambda}(t')\rp^2}{(t-t')^2} dt dt'\,.
   \ee
We decompose $A=a_0+z+z^\dagger$ with respect to 
the right and left channels,
so that $\lb \sigma_3, a_0\rb =0$, 
$\lb \sigma_3, z\rb =-2 z$,
$\lb \sigma_3, z^\dagger\rb =2 z^\dagger$. Then 
$A_\lambda\equiv
e^{-i\frac{\lambda}4\sigma_3} A e^{i\frac{\lambda}4\sigma_3}
=a_0+e^{i\frac{\lambda}2}z^\dagger+e^{-i\frac{\lambda}2}z$,
$B_\lambda=\lp e^{i\frac{\lambda}2}-e^{-i\frac{\lambda}2} \rp
W$, $W\equiv z^\dagger-z$.
Substituting this into (\ref{int1}) and (\ref{int2}) one rewrites
Eqs.(\ref{int1}),(\ref{int2}) 
in terms of $W(t)$:\footnotemark{} 
\footnotetext{Eq.(\ref{term1}) is essentially identical 
to the result obtained by 
Brouwer for the average pumped current\,\cite{Brouwer98}. The integral
in (\ref{term1}) is invariant under reparameterization, and thus 
has a purely geometric character determined by the contour $S(t)$
in $U(m)$.} 
  \be
\label{term1}
\frac{\sin\lambda}{8\pi}\oint {\rm tr} \lp \lb \sigma_3,W\rb \partial_t W\rp dt
  \ee
and
  \be
\label{term2}
\frac{\lp 1-\cos\lambda\rp}{2(2\pi)^2}\oint\!\!\oint 
\frac{{\rm tr} \lp W(t)-W(t')\rp^2}{(t-t')^2} dt dt' \,.
  \ee
From that we obtain the counting distribution for one pumping cycle:
\be\label{eq:chi-uv}
\chi(\lambda)=\exp\lp u(e^{i\lambda}-1)+v(e^{-i\lambda}-1)\rp
\ee
with the transmitted charge 
average $I=e(u-v)$ and  variance  
$J=e^2(u+v)$, related to the noise spectral 
density by ${\cal S}_0=J\Omega/\pi$ (see Sec.\ref{sec:tunneling}).

The parameters $u$ and $v$ in (\ref{eq:chi-uv}) can be expressed 
through $z(t)$ and $z^\dagger(t)$ as follows. 
Let us write $z(t)$ as $z_{+}(t)+z_{-}(t)$,
where $z_{+}(t)$ and $z_{-}(t)$ contain only 
positive or negative Fourier harmonics, respectively. Then
  \bea
\label{u-v}
u=\frac{i}{4\pi}\oint {\rm tr}\lp 
z^\dagger_{-}\partial_t z_{+} -
z_{+} \partial_t z^\dagger_{-}
\rp dt
=\sum\limits_{\omega>0} \omega\ {\rm tr} z^\dagger_{-\omega} z_{\omega}
\,,\\
v=\frac{i}{4\pi}\oint {\rm tr}\lp z_{-} \partial_t z^\dagger_{+}
-  z^\dagger_{+}\partial_t z_{-}
\rp dt
=-\sum\limits_{\omega<0} \omega\ {\rm tr} z^\dagger_{-\omega} z_{\omega}\,.
  \eea
This demonstartes that $u\ge0$ and $v\ge0$.
It is straightforward to show that (\ref{term1}) equals $i\sin\lambda (u-v)$,
whereas (\ref{term2}) equals $(\cos\lambda-1) (u+v)$.

Now we consider a single channel pump, $S(t)\in U(2)$. In this case, 
$z=z(t)\sigma_-$ and $z^\dagger=z^\ast(t)\sigma_+$.
For a harmonic driving signal, without loss of generality, one can write
  \be
  \label{V12}
z(t)=z_1V_1\cos(\Omega t+\theta)+z_2V_2\cos(\Omega t)
,
  \ee
where $V_{1,2}$ are pumping signal amplitudes, and complex parameters 
$z_{1,2}$ depend on microscopic details. From (\ref{u-v}) we find 
the rates
  \be
  \label{uv-2channel}
u=\frac14 \left| z_1V_1e^{i\theta}+z_2V_2 \right|^2 ,
\quad 
v=\frac14 \left| z_1V_1e^{-i\theta}+z_2V_2 \right|^2 .
  \ee
Both $u$ and $v$ can vanish at a particular amplitude 
ratio $V_1/V_2$ and phase $\theta$. When this happens, 
the two Poisson processes
(\ref{eq:chi-uv}) are reduced to one, and the current-to-noise ratio
gives elementary charge, $I/J=\pm e^{-1}$. This happens at
the extrema of $I/J$ as a function of $w=(V_1/V_2)e^{i\theta}$, 
for (\ref{uv-2channel}) reached
at $w=-z_2/z_1,\ -\bar z_2/\bar z_1$.

Reducing the counting statistics (\ref{eq:chi-uv}) to purely poissonian 
by varying pumping parameters
is possible, in principle, for any number of channels $n$.
However, since the number of parameters to be tuned is $2 n^2$, 
this method is practical perhaps  only for small channel numbers.
Although the method is demonstrated for non-interacting 
fermions, we argue that it can be applied to
interacting systems as well. Poisson statistics results from the absence 
of correlations of transmitted particles, which must be the 
case in a generic system, interacting or noninteracting, 
at small pumping current. Using the dependence of  
the rates $u$, $v$ on the driving signal to maximize $I/J$, i.e.
to eliminate one of the two Poisson processes (\ref{eq:chi-uv}), 
one could then 
obtain the charge quantum in the standard way as $e=J/I$.

To summarize the results of this section, in the weak pumping regime
the distribution of charge transmitted per cycle is of bidirectional 
Poisson form (\ref{eq:chi-bidirectional}), i.e., it 
is fully characterized by only two parameters, average current
and noise. The current to noise ratio $I/J$, scaled by elementary charge, 
varies between $1$ and $-1$, depending on the relation between driving signals
phases and amplitudes. 
Thus the quantity ${\rm max\/}(|I|/J)$ 
gives the inverse of elementary charge 
{\it without any fitting parameters\/}. Polianski et al.\,\cite{polianski02}
recently studied the dependence of $I/J$ on the mesoscopic scattering 
ensemble parameters, and found that, within the random matrix theory, 
the nearly extremal values close to $\pm 1$ can be reached 
with finite probability. This may permit
to use the noise in a pump to measure quasiparticle charge in open systems,
such as Luttinger liquids.

\end{document}